\documentclass[runningheads]{llncs}
\usepackage{graphicx}
\usepackage{amsmath}
\usepackage{mathtools}
\usepackage{algorithm}
\usepackage{algpseudocode}
\usepackage{tabularx}
\usepackage{booktabs}
\usepackage[T1]{fontenc}
\begin{document}
\title{Graph Enhanced Reinforcement Learning for Effective Group Formation in Collaborative Problem Solving}
%
%
\author{Zheng Fang\inst{1}\orcidID{0000-0001-9856-109X} \and
Fucai Ke \inst{2}\orcidID{0000-0001-9709-1305} \and Jae Young Han \inst{4} \orcidID{0000-0002-7701-9872} \and 
Zhijie Feng \inst{3}\orcidID{0009-0006-7980-2373} \and Toby Cai \inst{4}\orcidID{0009-0003-0690-1686}}

\institute{Centre for Learning Analytics, Monash University, Australia \\
\email{jardine.fang@monash.edu} \\ \and
Department of Human Centred Computing, Monash University, Australia \\
\email{fucai.ke1@monash.edu} \\ \and
Faculty of Pharmacy and Pharmaceutical Sciences, Monash University, Australia \\
\email{Jaeyoung.Han@monash.edu} \\ \and
University of Melbourne, Australia \\ 
\email{zhijief1@student.unimelb.edu.au} \\ \and
Meta Toby Tech Marketing \\
\email{digit.toby@gmail.com}}
%
\maketitle              
\begin{abstract}
This study addresses the challenge of forming effective groups in collaborative problem-solving environments. Recognizing the complexity of human interactions and the necessity for efficient collaboration, we propose a novel approach leveraging graph theory and reinforcement learning. Our methodology involves constructing a graph from a dataset where nodes represent participants, and edges signify the interactions between them. We conceptualize each participant as an agent within a reinforcement learning framework, aiming to learn an optimal graph structure that reflects effective group dynamics. Clustering techniques are employed to delineate clear group structures based on the learned graph. Our approach provides theoretical solutions based on evaluation metrics and graph measurements, offering insights into potential improvements in group effectiveness and reductions in conflict incidences. This research contributes to the fields of collaborative work and educational psychology by presenting a data-driven, analytical approach to group formation. It has practical implications for organizational team building, classroom settings, and any collaborative scenario where group dynamics are crucial. The study opens new avenues for exploring the application of graph theory and reinforcement learning in social and behavioral sciences, highlighting the potential for empirical validation in future work.

\keywords{Computational collaborative learning \and Graph Theory \and Reinforcement learning}
\end{abstract}

\section{Introduction}
    In the realm of collaborative problem solving, the formation of groups plays a pivotal role in determining the effectiveness and efficiency of the collaborative process \cite{nelson1999collaborative,graesser2018advancing,nokes2012effect,chang2017analysis}. Traditional methods of group formation often rely on intuition or simplistic criteria, which may not adequately account for the complex dynamics of human interactions. The need for a more systematic and data-driven approach is evident, particularly in environments where the stakes of collaboration are high, such as in organizational settings or educational contexts \cite{paquette2018matching,rahman2019optimized}.

    Recent advancements in graph theory and machine learning have opened up new possibilities for analyzing and understanding complex networks, including social and organizational structures \cite{rudin2014machine,ke2023hitskt,martin2023non,chen2020review,xiong2017deeppath,fang2021climate}. Graph theory, with its ability to model relationships and interactions between individuals, provides a robust framework for examining the underlying structure of collaborative groups. Meanwhile, reinforcement learning, a branch of machine learning, offers a powerful tool for optimizing complex systems, adapting to dynamic environments, and learning from interactions \cite{liu2022dynamic,wang2020grl,tiwari2021dapath,fang2021minimum}.

    The research utilized a dataset obtained from ``RescuShell"---a simulated internship program spanning 10 weeks, tailored for students in their first year of an undergraduate engineering course. The program cast students in the role of interns, where they were assigned the project of designing robotic legs for exoskeletons, a technology aimed at aiding rescue operations. The participant pool consisted of forty-eight students, and the focus of the analysis was on the initial eleven tasks out of the program's total duration. During this phase, the students were distributed into ten distinct groups, with each group being responsible for exploring different types of actuators that could potentially be integrated into the exoskeleton design. This study utilized thematic codes that were established in prior research \cite{siebert2017search}.
    
    This study bridges the fields of graph theory and reinforcement learning to tackle the challenge of optimal group formation in collaborative problem-solving scenarios. The rationale for mapping collaborative learning cases onto a graph structure lies in the inherent network-like nature of social interactions. In such a graph, nodes represent individual participants, effectively capturing the discrete entities involved in collaboration. Edges, with associated weights, represent the relationships and levels of interaction between these participants, mirroring the complex web of interpersonal dynamics that characterize collaborative efforts.

    By treating each participant as an agent within a reinforcement learning framework, we not only reflect the individuality of participants but also their capacity to adapt and learn within a group. This approach allows for the exploration of various configurations and interactions, with the goal of identifying an optimal graph structure that fosters effective group dynamics. The participants, modeled as agents, learn to navigate and adjust within this network, simulating real-world adaptations and strategic interactions within a group.
    
    Following the reinforcement learning phase, we employ clustering techniques to group participants. This step is crucial in translating the theoretical graph structure into practical group formations. The clustering is designed to aggregate suitable individuals into cohesive groups, respecting a predetermined group size. 
    
    Our approach is innovative in its application of reinforcement learning to graph-theoretic models in a social context. While previous research has separately explored graph theory in social networks and reinforcement learning in optimization problems, their combination in the context of group formation is relatively unexplored. This study aims to fill this gap, providing a novel perspective on how data-driven, analytical methods can enhance our understanding and practice of forming collaborative groups.
    
    The methodology we propose goes beyond traditional approaches by not just forming groups, but by also learning from and adapting to the intricate patterns of human interaction. This adaptive quality is particularly relevant in today's rapidly changing social and organizational environments, where flexibility and responsiveness are key to successful collaboration.
    
    In summary, our study contributes to the fields of social network analysis, organizational behavior, and educational psychology, offering a new lens through which to view the perennial challenge of effective group formation. 

\section{Background}
    The concept of collaborative problem solving has been extensively studied in both social and educational settings \cite{nelson1999collaborative,graesser2018advancing,hofmann2015overcoming}. Research in this area focuses on how individuals interact within a group to achieve common goals, emphasizing the importance of effective communication, conflict resolution, and collective decision-making. Studies have shown that well-structured groups can lead to higher levels of engagement, learning, and problem-solving efficiency. However, the challenge of forming these optimal groups often relies on traditional methods that may not consistently yield the best results.

    Traditionally, group formation in collaborative learning has been guided by strategies such as random assignment, self-selection, or instructor-based grouping \cite{graesser2020collaboration,roschelle1995construction,fang2023automated,fang2024neural}. Random assignment, while easy to implement and ensuring a degree of fairness, does not take into account the diverse characteristics and compatibility of participants. Self-selection allows participants to choose their own groups, which can lead to comfort and ease of collaboration, but often results in homogeneity and the potential exclusion of less popular individuals. Instructor-based grouping, wherein an educator forms groups based on their perception of student skills and personalities, attempts to balance group dynamics but can be limited by the instructor’s subjective judgment and lack of detailed insight into every student’s capabilities and interaction patterns \cite{chen2019optimized,ding2009visualizing}.
    
    These traditional methods, while practical, often overlook the complex underlying dynamics that influence the effectiveness of a collaborative group. Factors such as interpersonal relations, individual strengths and weaknesses, and the specific nature of the task at hand are seldom fully considered. This can lead to suboptimal group compositions, where the potential for conflict is higher and the synergy needed for effective collaboration is lacking \cite{zambrano2019effects,stasser2020collective}. Consequently, there is a growing recognition of the need for more sophisticated, data-driven approaches that can analyze and optimize group dynamics based on a comprehensive understanding of individual and collective attributes.

    Graph theory has emerged as a powerful tool in understanding the complexities of social networks. By representing individuals as nodes and their relationships as edges, researchers can analyze patterns of interaction and influence within a group \cite{wu2020comprehensive,chen2020review}. This approach has been instrumental in identifying key actors, understanding group dynamics, and predicting the flow of information in various social settings. The application of graph theory in collaborative group formation is a relatively new but promising area of study.
    
    Reinforcement learning (RL) is a type of machine learning where agents learn to make decisions by interacting with an environment \cite{wiering2012reinforcement,franccois2018introduction}. In the context of group dynamics, RL can be used to optimize group structures based on certain performance metrics. While its application in social sciences is still emerging, RL has shown potential in developing adaptive systems that can learn from and respond to complex interactions.
    
    The integration of graph theory and RL in the context of group formation is a novel approach. While there have been studies using graph-theoretic models to understand social networks and others using RL for optimization, the combined use of these two methods to optimize group formation in collaborative settings is innovative. This approach allows for a more nuanced and dynamic understanding of group interactions and offers a data-driven method to form effective groups.
    
    Despite the theoretical advancements in using graph theory and RL in social sciences, there remains a gap in empirical validation, particularly in real-world group formation scenarios. Studies have often been limited to simulations or controlled experiments, and there is a need for research that tests these models in actual collaborative environments.

\section{Methodology}
    \subsection{Graph Formation from Dataset}
        In this section, we describe the methodology for constructing a graph from the input data, which serves as the foundational structure for our analysis. The graph is formed by representing participants as nodes and their interactions as weighted edges.
        
        \subsubsection{Data Preprocessing and Node Representation}
            The dataset, comprising various attributes such as participant names, group affiliations, and thematic codes, undergoes a preprocessing stage. Participant and group names are standardized, ensuring uniformity and consistency in representation. Subsequently, each unique participant is assigned a distinct numerical identifier, denoted by $u_i$ where $i$ is the index of the participant. Similarly, each group is assigned a unique identifier $g_j$.

        \subsubsection{Graph Construction and Edge Formulation}
            We construct an undirected graph $G = (U, E)$, where $U$ is the set of nodes representing participants, and $E$ is the set of edges representing interactions between them. Each edge $e_{ij}$ in the graph corresponds to an interaction between participants $u_i$ and $u_j$. The weight of an edge $w_{ij}$ is determined by the extent of interaction between these participants, quantified based on thematic codes.

        \subsubsection{Edge Weight Calculation}
            For each pair of participants in the same group, we analyze the thematic codes to identify shared attributes. If a thematic code is attributed to both participants, the weight of the edge connecting them, $w_{ij}$, is incremented, symbolizing the strength of their interaction. The weight of each edge is normalized by dividing it by the maximum weight in the graph, ensuring that all weights fall within a standard range. This normalization is represented by the Equation \ref{edgenormalization}.
            \begin{equation}
            \label{edgenormalization}
                w_{ij}' = \frac{w_{ij}}{\displaystyle\max_{(u_i,u_j) \in E}(w_{ij})}
            \end{equation}

    \subsection{Reinforcement Learning Environment and Reward Functions}
        \subsubsection{Environment Setup}
            The reinforcement learning environment is designed around an undirected graph representing the collaborative network. This environment, encapsulating the interactions and relationships of participants, serves as the basis for our optimization problem. The environment is characterized by a graph with a predefined set of nodes and edges, where each node represents a participant and each edge denotes the interaction between participants. The learning agent interacts with this environment with the goal of optimizing the graph structure to enhance collaborative efficacy. The reinforcement learning agent's objective is to modify the graph structure in a manner that maximizes the composite reward, which is a balanced amalgamation of these metrics. Each action taken by the agent alters the graph, influencing the subsequent state and the associated reward. The learning process continues iteratively, with the agent exploring various modifications to the graph structure to identify the configuration that yields the highest reward.

        \subsubsection{Reward Functions}
            The optimization process is driven by a composite reward function composed of several key metrics:
            \begin{itemize}
                \item \textbf{Maximized Overall Connectivity:} This metric evaluates the total weight of the graph's edges relative to the maximum possible total weight. It is given by the formula:
                \begin{equation}
                    OC = \dfrac{\sum\limits_{e_{ij} \in E} w_{ij}}{\dfrac{n(n - 1)}{2}} 
                \end{equation}
                
                where \( \sum_{e_{ij} \in E} w_{ij} \) is the sum of the weights of all edges in the graph, \( n \) is the number of nodes, and \( \frac{n(n - 1)}{2} \) represents the maximum number of edges in an undirected graph.
                
                \item \textbf{Balanced Participation:} The variance in the degrees of nodes within the graph is evaluated to ensure equitable involvement. The formula for variance is:
                \begin{equation}
                    \mathrm{Var} = \frac{1}{n}\sum_{i=1}^{n}\left(d_i - \bar{d}\right)^2
                \end{equation}
                where \( d_i \) is the degree of the \( i^{\text{th}} \) node, and \( \bar{d} \) is the average degree of nodes in the graph.
                
                \item \textbf{Efficiency of Communication:} This criterion is quantified by calculating the average path length between nodes, formulated as:
                \begin{equation}
                    \mathrm{P}_l = \frac{1}{\frac{n(n - 1)}{2}} \sum_{i=1}^{n}\sum_{j=i+1}^{n} l_{ij}
                \end{equation}
                where \( l_{ij} \) is the length of the shortest path between nodes \( i \) and \( j \), and \( \frac{n(n - 1)}{2} \) is the total number of node pairs.
                
                \item \textbf{Penalty for Dominance:} To discourage dominance, a penalty based on the disparity between the highest degree of a node and the average degree is used:
                \begin{equation}
                     \mathrm{P}_d = \max(d_i) - \bar{d}
                \end{equation}
                where \( \max(d_i) \) is the highest degree among all nodes, and \( \bar{d} \) is the average degree of nodes in the graph.
            \end{itemize}

    \subsection{Graph Update in Reinforcement Learning Environment}
        The reinforcement learning environment allows for the dynamic modification of the graph structure based on the actions taken by the learning agent \cite{xiong2017deeppath}. The process of updating the graph is a key component of our methodology, enabling the exploration of different configurations and their impact on group dynamics.

        \subsubsection{Action Application and Graph Modification}
            Each action taken by the agent is interpreted as an instruction to modify the graph. The actions are continuous values, normalized to the range [0, 1], representing the intensity or probability of the modification. Formally, an action $\alpha$ is transformed to a normalized action $a'$ using the transformation $a' = \frac{\alpha + 1}{2}$, where $\alpha \in [-1, 1]$.

        \subsubsection{Probabilistic Edge Selection}
            The agent selects an edge within the graph to modify based on the normalized action. The selection process is probabilistic, with the probability of selecting an edge $e_{ij}$ between nodes $u_i$ and $u_j$ influenced by the current edge weight $w_{ij}$ and the action probability $a'$. The adjusted probability $P(e_{ij})$ for selecting an edge is computed as follows:
            \begin{equation}
            \label{edgeselection}
                P(e_{ij}) = \dfrac{w_{ij} \cdot a'}{\sum\limits_{e_{kl} \in E}(w_{kl} \cdot a')}
            \end{equation}
            where $E$ is the set of all edges in the graph.

        \subsubsection{Edge Weight Modification}
            Upon selecting an edge, the weight of the edge is modified in accordance with the action probability. The new weight $w'_{ij}$ of the edge $e_{ij}$ is calculated by adding the normalized action value to the current weight $w'_{ij} = w_{ij} + \alpha'$

        \subsubsection{Environment State Update and Reward Calculation}
            After applying the action and modifying the graph, the environment state is updated. The reward is calculated based on the modified graph structure, taking into account the overall connectivity, balanced participation, efficiency of communication, and penalty over dominance. This reward guides the learning process, encouraging the agent to explore graph configurations that optimize these criteria. The learning agent iteratively applies actions to modify the graph, with each step consisting of an action application, graph update, and reward calculation. The process continues until a termination condition is met, such as reaching a maximum number of iterations or achieving a predefined optimization level.

    \subsection{Multi-Agent Deep Deterministic Policy Gradient (MADDPG) Framework}
        In the Multi-Agent Deep Deterministic Policy Gradient (MADDPG) framework, the architecture of the agents is a cornerstone for learning optimal actions within the collaborative network \cite{li2019robust}. Each agent comprises two neural network components: the Actor network and the Critic network. These networks are designed to interact with each other, enabling agents to make informed decisions and evaluate the potential outcomes of these decisions.
        \subsubsection{Actor Network}
            The Actor network is responsible for mapping the state of the environment to an action. Its architecture consists of a series of fully connected layers, designed to process the input state and output the corresponding action. Specifically, the network has the following structure:
            \begin{algorithm}
            \caption{Actor Network Architecture}
            \begin{algorithmic}[1]
            \State \textbf{Input:} state\_size, action\_size, hidden\_sizes
            \Function{Forward}{state}
                \State $x \gets \mathrm{ReLU}\mathopen{}(fc1\mathopen{}(\text{state}))$
                \State $x \gets \mathrm{ReLU}\mathopen{}(fc2\mathopen{}(x))$
                \State \Return $\tanh(fc3\mathopen{}(x))$ \Comment{Action space assumed to be between $-1$ and $1$}
            \EndFunction
            \end{algorithmic}
            \end{algorithm}

        \subsubsection{Critic Network}
            The Critic network evaluates the potential reward of the action proposed by the Actor network, given the current state of the environment. It is structured similarly to the Actor network but with a key difference in the input and output layers:
            \begin{algorithm}
            \caption{Critic Network Architecture}
            \begin{algorithmic}[1]
            \State \textbf{Input:} state\_size, action\_size, hidden\_sizes
            \Function{Forward}{state, action}
                \State $x \gets \mathrm{ReLU}\mathopen{}(fc1\mathopen{}(\text{Concatenate}(\text{state, action})))$
                \State $x \gets \mathrm{ReLU}\mathopen{}(fc2\mathopen{}(x))$
                \State \Return $fc3(x)$
            \EndFunction
            \end{algorithmic}
            \end{algorithm}

            Both networks are integral to the reinforcement learning process. The Actor network's role is to determine the best possible action in a given state, while the Critic network assesses the quality of these actions, guiding the Actor towards more beneficial decisions. Through this collaborative process, agents in the MADDPG framework can effectively learn and adapt their strategies to optimize group formations in the collaborative problem-solving environment.

        \subsubsection{Noise Process}
            In the MADDPG framework, two crucial components for efficient learning are the noise process and the replay memory. These components play pivotal roles in exploration and experience replay, respectively. The framework employs an Ornstein-Uhlenbeck process to add noise to the actions. This type of noise is particularly useful in physical control problems with inertia, as it generates temporally correlated values, providing a more realistic variation in actions.
            \begin{algorithm}
            \caption{Noise Process in MADDPG}
            \begin{algorithmic}[1]
            \State \textbf{Ornstein-Uhlenbeck Noise Process:}
            \begin{itemize}
                \item Initialized with parameters: mean $\mu$, speed of mean reversion $\theta$, and noise scale $\sigma$.
                \item At each time step, noise is updated as:
                \[ \Delta x = \theta (\mu - x) + \sigma \mathcal{N}\mathopen{}(0,1) \]
                \item The noise process is reset at the start of each episode to $\mu$.
            \end{itemize}
            \end{algorithmic}
            \end{algorithm}

        \subsubsection{Replay Memory}
            Replay memory is a critical component that enables the agents to learn from past experiences. It stores a collection of experiences, each comprising a state, action, reward, next state, and a ``done" flag indicating whether the episode has ended.
            \begin{algorithm}
            \caption{Replay Memory in MADDPG}
            \begin{algorithmic}[1]
            \State \textbf{Replay Memory:}
            \begin{itemize}
                \item Stores experiences as tuples \textit{(state, action, reward, next\_state, done)}.
                \item Fixed-size buffer; older experiences are discarded when the limit is reached.
                \item Experiences are randomly sampled for learning, providing diverse learning instances.
                \item Preprocessing of states and next states for network compatibility.
            \end{itemize}
            \end{algorithmic}
            \end{algorithm}

            When the agent decides to learn, a batch of experiences is sampled from the replay memory. The learning process then uses this batch to update the Actor and Critic networks, refining the policy and value estimations based on a diverse range of past experiences. This experience replay mechanism is a key feature of deep reinforcement learning, enabling efficient and robust learning from past interactions with the environment.

    \subsection{Clustering for Group Formation}
        Following the reinforcement learning phase, we employ a clustering algorithm to form optimal groups based on the learned graph structure. This step is crucial in translating the theoretical graph structure into practical group formations. Our clustering approach, termed ``constrained clustering", aims to distribute participants into a predefined number of clusters, each with a specific size limit \cite{rose1993constrained}. This method ensures that groups are evenly sized and manageable, adhering to constraints that are often present in real-world collaborative settings.

        The process begins with the list of participants, each represented as a node in the previously optimized graph. These participants are randomly shuffled to prevent any initial ordering bias. We then establish a set number of clusters, each initially empty, and a corresponding array to track the number of participants in each cluster.
        
        The algorithm iteratively assigns each participant to one of the clusters. The assignment is based on the current occupancy of the clusters, with priority given to those with the fewest members \cite{basu2008constrained}. This strategy ensures a balanced distribution of participants across the clusters. Once a participant is assigned to a cluster, the count for that cluster is incremented. If a cluster reaches its predetermined size limit, it is no longer considered for further assignments. This constraint maintains the uniformity in the size of the groups.
        
        The clustering continues until all participants are assigned to a cluster. The result is a set of optimally formed groups, each consisting of a specific number of participants. These groups are expected to exhibit balanced participation and efficient collaboration, as indicated by the reinforcement learning phase and the clustering algorithm.

\section{Results}
\begin{figure}[h]
        \centering
        \includegraphics[width=\linewidth, height=8cm]{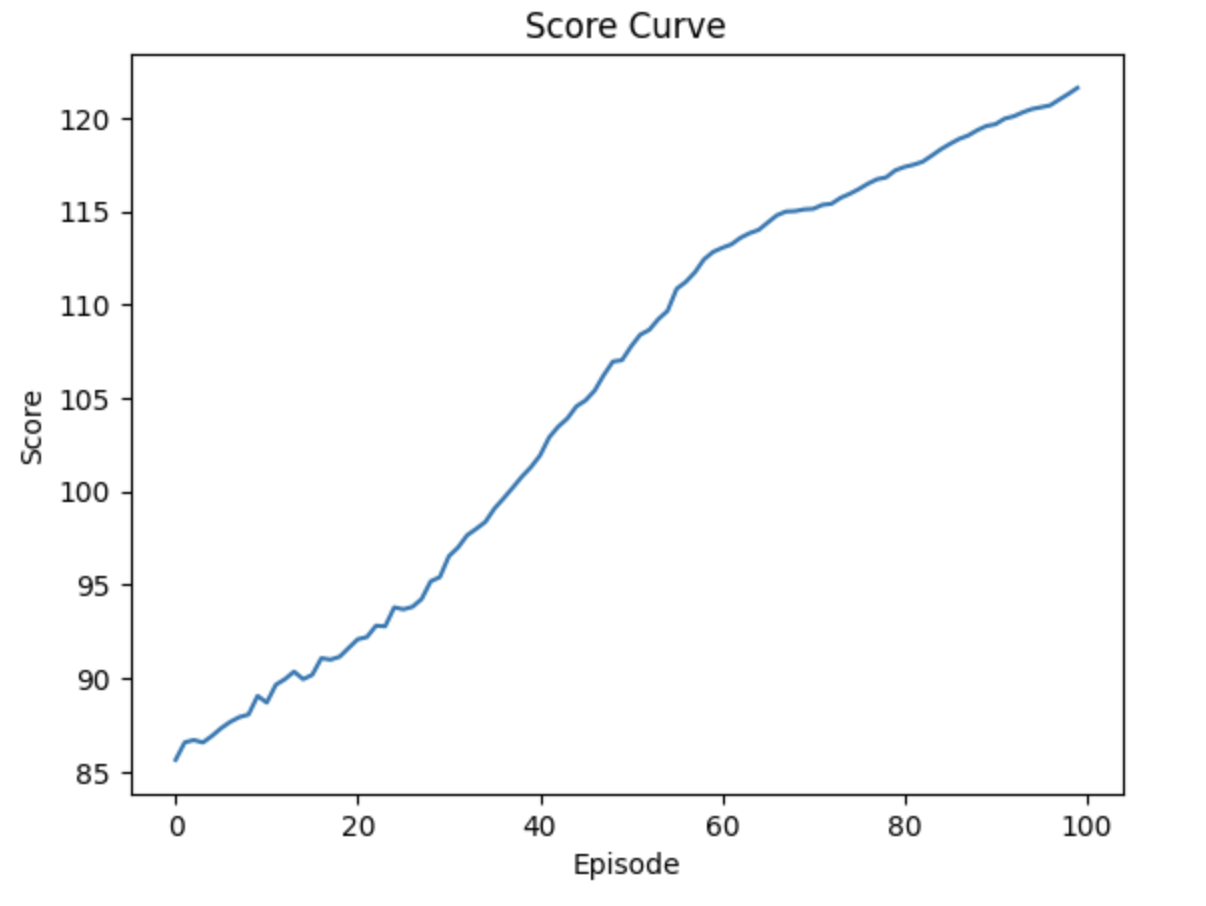}  
        \caption{Scores from reward functions}
        \label{scores}
    \end{figure}

    The image \ref{scores} appears to show a positive trend, with scores increasing over the index, which likely corresponds to either episodes or iterations of the reinforcement learning process. The initial scores start off lower, which is expected as the learning agents begin by exploring the environment and understanding the dynamics of the graph structure. As learning progresses, we can infer that the agents improve their policy, as evidenced by the increasing trend in the scores. This suggests that the agents are effectively learning to maximize the rewards over time, which could be due to the refinement of their strategies and the optimization of the graph structure for better group dynamics.

    Towards the latter part of the curve, the slope seems to decrease slightly, indicating that the rate of improvement in scores is stabilizing. This could be a sign that the agents are approaching an optimal policy, where further improvements are marginal. The absence of significant dips in the score curve suggests a stable learning process without drastic policy degradations. Overall, the increasing trend is indicative of successful learning, with the reinforcement learning algorithm progressively enhancing the collaborative network's performance as per the defined reward functions.

\begin{figure}[!htbp]
        \centering
        \includegraphics[width=\linewidth]{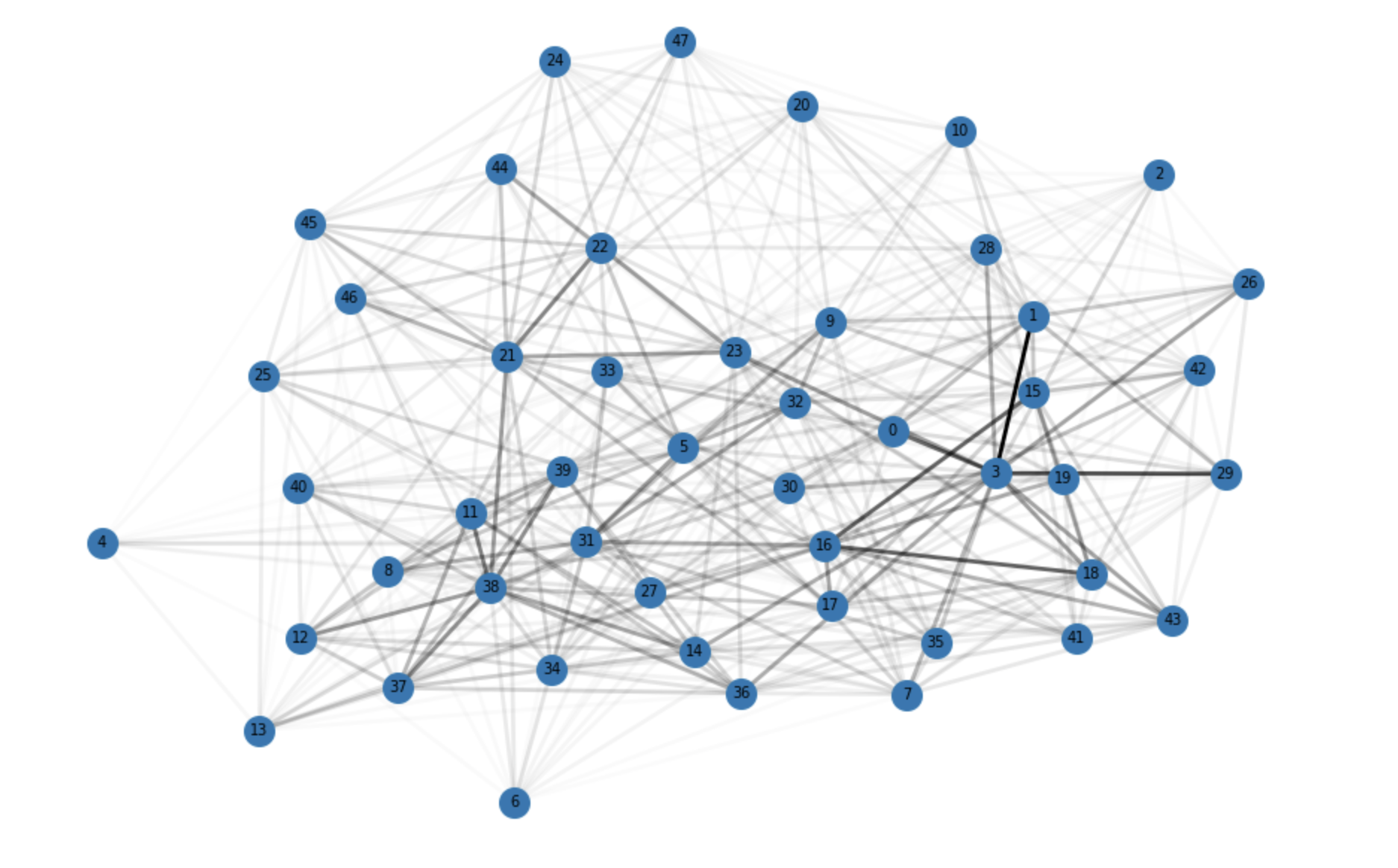}  
        \caption{Original Graph}
        \label{fig:kg}
    \end{figure}

\begin{figure}[!htbp]
        \centering
        \includegraphics[width=\linewidth]{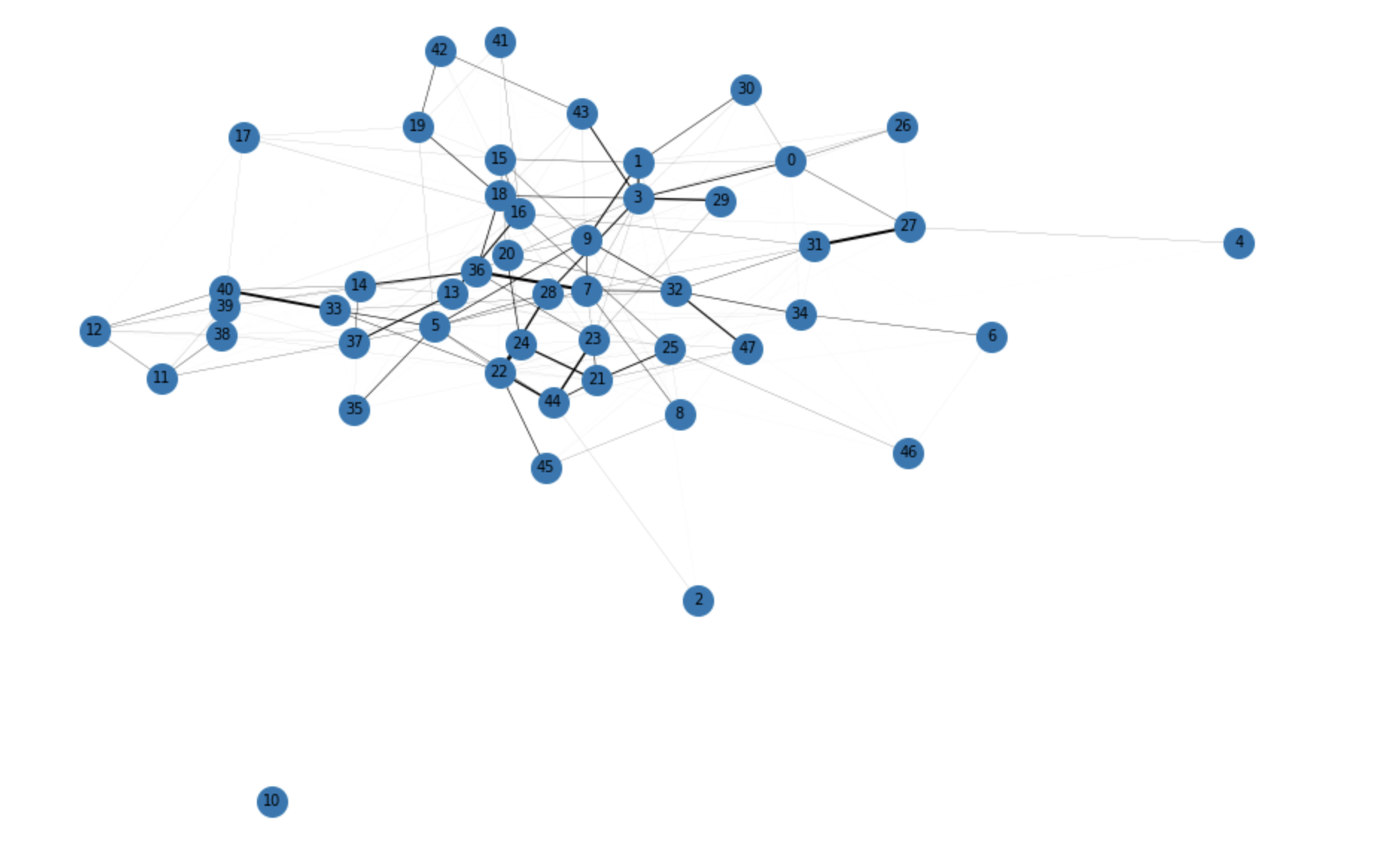}  
        \caption{Graph from reinforcement learning}
        \label{fig:rl}
    \end{figure}

    The transition from the initial graph \ref{fig:kg} structure to the one refined by reinforcement learning \ref{fig:rl} marks significant changes in the collaborative network's dynamics. By examining the two images – the original graph constructed from the dataset and the graph post-reinforcement learning – we can deduce the impact of our optimization process.

    In the initial graph, we observe a dense network with numerous overlapping edges and a relatively uniform distribution of connections. This representation indicates a high degree of interconnectedness among participants; however, such density can also imply potential redundancies in communication pathways and a lack of clear subgroup structures.
    
    Post-reinforcement learning, the graph presents a different picture. The node connections appear to be more selective, with a reduction in the overall edge density. This suggests that the reinforcement learning algorithm has optimized the network by possibly strengthening relevant connections while pruning less useful ones, thereby potentially enhancing the efficiency of the collaborative network. Moreover, the structure now shows a more discernible pattern of subgroups or clusters, which may indicate the emergence of specialized groups within the larger network.
    
    The reinforced graph likely represents a more streamlined communication structure, where the reduced complexity could translate into faster and more effective information transfer among participants. The formation of subgroups might reflect the algorithm's strategy to optimize collaboration by creating teams with complementary skills or aligned objectives.
    
    The visual differences between the two graphs underscore the reinforcement learning algorithm's effectiveness in restructuring the collaborative network towards a potentially more functional and efficient configuration. These observed changes provide a basis for the subsequent clustering process, which aims to formalize these naturally emergent subgroups into distinct collaborative teams.

    The table \ref{tab:clustering_results} presents the distribution of participant IDs across ten distinct groups, resulting from the constrained clustering algorithm. Groups vary in size, with containing four or five participants. The clustering results suggest a strategy that seeks to balance the distribution of participants while possibly considering the strength and quality of connections between them, as informed by the reinforcement learning phase. For instance, the larger group may comprise individuals who share stronger or more numerous connections, indicating a central hub of activity or a sub-network within the larger collaborative network. Conversely, the smaller groups suggest more specialized or tightly-knit collaborations. This formation of groups is to facilitate efficient collaboration, with the potential for both broad-based and intensive problem-solving depending on the group's composition.

    \begin{table}[htbp!]
        \centering
        \caption{Clustering Results}
        \begin{tabularx}{\textwidth}{>{\centering}X>{\arraybackslash}X}
        \toprule
        \textbf{Group} & \textbf{Participant IDs} \\ \midrule
        0              & 26, 1, 9, 28, 39                            \\
        1              & 22, 25, 14, 7, 3, 40, 15                    \\
        2              & 11, 0, 5, 16                                \\
        3              & 37, 20, 18, 31, 32                          \\
        4              & 24, 47, 6, 21                               \\
        5              & 2, 38, 12, 46, 27                           \\
        6              & 17, 44, 29, 45, 35                          \\
        7              & 36, 10, 23, 8                               \\
        8              & 13, 4, 19, 41                               \\
        9              & 33, 43, 30, 34, 42                          \\ 
        \bottomrule
        \end{tabularx}
        \label{tab:clustering_results}
    \end{table}

\section{Discussion}
    In educational settings, the representation of learners as nodes in a graph aligns with Vygotsky's social development theory which posits that social interaction plays a fundamental role in the development of cognition \cite{goldstein2011vygotskys}. By mapping participants to nodes and their interactions to edges, we encapsulate the essence of collaborative learning—each individual's learning experience is interlinked with that of their peers, and knowledge is constructed through these connections.

    The optimization of this graph via reinforcement learning further adheres to the principles of collaborative learning. As each participant or agent seeks to maximize their reward through interactions with others, the system dynamically evolves towards optimized learning pathways and group formations. This mirrors the adaptive nature of human learning, where individuals continuously adjust their learning strategies in response to the feedback from their environment. From an educational perspective, the reinforcement learning process can be viewed as a metaphor for formative assessment, where ongoing feedback is used to guide and improve learners' performance and interaction patterns.
    
    Moreover, the concept of optimizing group formations based on graph theory and reinforcement learning has practical implications in real-life educational settings. It suggests a data-driven approach to group formation that could lead to more effective learning outcomes, as groups are formed based on empirical evidence of interaction patterns and performance metrics rather than intuition. This is in line with contemporary educational models that emphasize personalized learning and adaptive teaching strategies. By forming groups that are theoretically optimized for reduced conflict and enhanced collaboration, educators can potentially foster a more conducive learning environment that caters to the diverse needs and strengths of learners. 

    However, we acknowledge certain limitations in our current approach. Firstly, the lack of robust evaluation metrics means that the effectiveness of the group formations is not quantified against established benchmarks. Secondly, the absence of real human experiments to validate the theoretical model limits the empirical applicability of our findings. Without testing in a real-world environment, the practical implications of the optimized group structures remain speculative.

\section{Conclusion}
    This study embarked on an innovative exploration of group formation within collaborative problem-solving contexts, utilizing graph theory and reinforcement learning to model and optimize the interactions between participants. The application of a reinforcement learning algorithm to this graph structure allowed us to simulate an optimization process that reflects the adaptive and iterative nature of human learning and social interactions. To address limitations and further this line of research, future work will incorporate graph neural networks within the reinforcement learning model. This advancement aims to capture more nuanced interaction patterns and participant attributes, potentially leading to more sophisticated group optimization. Moreover, by integrating domain knowledge, we plan to construct knowledge graphs that embody the contextual intricacies of collaborative problem solving. This will enable the development of tailored evaluation metrics, facilitating quantitative assessments of group performance and the efficacy of optimized formations. The ultimate goal of future improvements is to align the theoretical model more closely with real-world educational and collaborative settings. Through empirical validation and the integration of domain-specific knowledge, we aspire to refine our approach to group formation, contributing valuable insights and tools to the fields of educational psychology and collaborative work.

\bibliographystyle{unsrt}
\bibliography{ref.bib}

\begin{thebibliography}{10}

\bibitem{nelson1999collaborative}
Laurie~Miller Nelson.
\newblock Collaborative {Problem} {Solving}.
\newblock In {\em Instructional-design {Theories} and {Models}}, pages 241--267. Routledge, 1999.

\bibitem{graesser2018advancing}
Arthur~C Graesser, Stephen~M Fiore, Samuel Greiff, Jessica Andrews-Todd, Peter~W Foltz, and Friedrich~W Hesse.
\newblock Advancing the science of collaborative problem solving.
\newblock {\em psychological science in the public interest}, 19(2):59--92, 2018.

\bibitem{nokes2012effect}
Timothy~J Nokes-Malach, Michelle~L Meade, and Daniel~G Morrow.
\newblock The effect of expertise on collaborative problem solving.
\newblock {\em Thinking \& Reasoning}, 18(1):32--58, 2012.

\bibitem{chang2017analysis}
Chia-Jung Chang, Ming-Hua Chang, Bing-Cheng Chiu, Chen-Chung Liu, Shih-Hsun~Fan Chiang, Cai-Ting Wen, Fu-Kwun Hwang, Ying-Tien Wu, Po-Yao Chao, Chia-Hsi Lai, et~al.
\newblock An analysis of student collaborative problem solving activities mediated by collaborative simulations.
\newblock {\em Computers \& Education}, 114:222--235, 2017.

\bibitem{paquette2018matching}
Luc Paquette, Nigel Bosch, Emma Mercier, Jiyoon Jung, Saadeddine Shehab, and Yurui Tong.
\newblock Matching data-driven models of group interactions to video analysis of collaborative problem solving on tablet computers.
\newblock In {\em International Conference of the Learning Sciences}, 2018.

\bibitem{rahman2019optimized}
Habibur Rahman, Senjuti~Basu Roy, Saravanan Thirumuruganathan, Sihem Amer-Yahia, and Gautam Das.
\newblock Optimized group formation for solving collaborative tasks.
\newblock {\em The VLDB Journal}, 28:1--23, 2019.

\bibitem{rudin2014machine}
Cynthia Rudin and Kiri~L Wagstaff.
\newblock Machine learning for science and society, 2014.

\bibitem{ke2023hitskt}
Fucai Ke, Weiqing Wang, Weicong Tan, Lan Du, Yuan Jin, Yujin Huang, and Hongzhi Yin.
\newblock Hitskt: A hierarchical transformer model for session-aware knowledge tracing.
\newblock {\em Knowledge-Based Systems}, page 111300, 2023.

\bibitem{martin2023non}
Cameron Martin, Fucai Ke, and Hao Wang.
\newblock Non-intrusive load monitoring for feeder-level {EV} charging detection: Sliding window-based approaches to offline and online detection, 2023.
\newblock arXiv:2312.01887 [cs, eess].

\bibitem{chen2020review}
Xiaojun Chen, Shengbin Jia, and Yang Xiang.
\newblock A review: Knowledge reasoning over knowledge graph.
\newblock {\em Expert Systems with Applications}, 141:112948, 2020.

\bibitem{xiong2017deeppath}
Wenhan Xiong, Thien Hoang, and William~Yang Wang.
\newblock {DeepPath}: A reinforcement learning method for knowledge graph reasoning, 2018.
\newblock arXiv:1707.06690 [cs].

\bibitem{fang2021climate}
Zheng Fang, Jianying Xie, Ruiming Peng, and Sheng Wang.
\newblock Climate finance: Mapping air pollution and finance market in time series.
\newblock {\em Econometrics}, 9(4):43, 2021.

\bibitem{liu2022dynamic}
Hao Liu, Shuwang Zhou, Changfang Chen, Tianlei Gao, Jiyong Xu, and Minglei Shu.
\newblock Dynamic knowledge graph reasoning based on deep reinforcement learning.
\newblock {\em Knowledge-Based Systems}, 241:108235, 2022.

\bibitem{wang2020grl}
Qi~Wang, Yuede Ji, Yongsheng Hao, and Jie Cao.
\newblock Grl: Knowledge graph completion with gan-based reinforcement learning.
\newblock {\em Knowledge-Based Systems}, 209:106421, 2020.

\bibitem{tiwari2021dapath}
Prayag Tiwari, Hongyin Zhu, and Hari~Mohan Pandey.
\newblock Dapath: Distance-aware knowledge graph reasoning based on deep reinforcement learning.
\newblock {\em Neural Networks}, 135:1--12, 2021.

\bibitem{fang2021minimum}
Zheng Fang, David~L Dowe, Shelton Peiris, and Dedi Rosadi.
\newblock Minimum message length in hybrid arma and lstm model forecasting.
\newblock {\em Entropy}, 23(12):1601, 2021.

\bibitem{siebert2017search}
Amanda~Lee Siebert-Evenstone, Golnaz~Arastoopour Irgens, Wesley Collier, Zachari Swiecki, Andrew~R Ruis, and David~Williamson Shaffer.
\newblock In search of conversational grain size: Modeling semantic structure using moving stanza windows.
\newblock {\em Journal of Learning Analytics}, 4(3):123--139, 2017.

\bibitem{hofmann2015overcoming}
David~A Hofmann.
\newblock Overcoming the obstacles to cross-functional decision making: laying the groundwork for collaborative problem solving.
\newblock {\em Organizational Dynamics}, 44(1):17--25, 2015.

\bibitem{graesser2020collaboration}
Arthur~C. Graesser, Samuel Greiff, Matthias Stadler, and Keith~T. Shubeck.
\newblock Collaboration in the 21st century: {The} theory, assessment, and teaching of collaborative problem solving.
\newblock {\em Computers in Human Behavior}, 104:106134, 2020.

\bibitem{roschelle1995construction}
Jeremy Roschelle and Stephanie~D. Teasley.
\newblock The construction of shared knowledge in collaborative problem solving.
\newblock In Claire O'Malley, editor, {\em Computer {Supported} {Collaborative} {Learning}}, {NATO} {ASI} {Series}, pages 69--97, Berlin, Heidelberg, 1995. Springer.

\bibitem{fang2023automated}
Zheng Fang, Ying Yang, and Zachari Swiecki.
\newblock Automated code discovery via graph neural networks and generative ai.
\newblock In {\em International Conference on Quantitative Ethnography}, pages 438--454. Springer, 2023.

\bibitem{fang2024neural}
Zheng Fang, Weiqing Wang, Guanliang Chen, and Zachari Swiecki.
\newblock Neural epistemic network analysis: Combining graph neural networks and epistemic network analysis to model collaborative processes.
\newblock In {\em Proceedings of the 14th Learning Analytics and Knowledge Conference}, pages 157--166, 2024.

\bibitem{chen2019optimized}
Chih-Ming Chen and Chi-Hsiung Kuo.
\newblock An optimized group formation scheme to promote collaborative problem-based learning.
\newblock {\em Computers \& Education}, 133:94--115, 2019.

\bibitem{ding2009visualizing}
Ning Ding.
\newblock Visualizing the sequential process of knowledge elaboration in computer-supported collaborative problem solving.
\newblock {\em Computers \& Education}, 52(2):509--519, 2009.

\bibitem{zambrano2019effects}
Jimmy Zambrano, Femke Kirschner, John Sweller, and Paul~A Kirschner.
\newblock Effects of group experience and information distribution on collaborative learning.
\newblock {\em Instructional Science}, 47:531--550, 2019.

\bibitem{stasser2020collective}
Garold Stasser and Susanne Abele.
\newblock Collective choice, collaboration, and communication.
\newblock {\em Annual Review of Psychology}, 71:589--612, 2020.

\bibitem{wu2020comprehensive}
Zonghan Wu, Shirui Pan, Fengwen Chen, Guodong Long, Chengqi Zhang, and S~Yu Philip.
\newblock A comprehensive survey on graph neural networks.
\newblock {\em IEEE transactions on neural networks and learning systems}, 32(1):4--24, 2020.

\bibitem{wiering2012reinforcement}
Marco~A Wiering and Martijn Van~Otterlo.
\newblock Reinforcement learning.
\newblock {\em Adaptation, learning, and optimization}, 12(3):729, 2012.

\bibitem{franccois2018introduction}
Vincent François-Lavet, Peter Henderson, Riashat Islam, Marc~G. Bellemare, and Joelle Pineau.
\newblock An {Introduction} to {Deep} {Reinforcement} {Learning}.
\newblock {\em Foundations and Trends{\textregistered} in Machine Learning}, 11(3-4):219--354, 2018.
\newblock Publisher: Now Publishers, Inc.

\bibitem{li2019robust}
Shihui Li, Yi~Wu, Xinyue Cui, Honghua Dong, Fei Fang, and Stuart Russell.
\newblock Robust multi-agent reinforcement learning via minimax deep deterministic policy gradient.
\newblock In {\em Proceedings of the {Thirty}-{Third} {AAAI} {Conference} on {Artificial} {Intelligence} and {Thirty}-{First} {Innovative} {Applications} of {Artificial} {Intelligence} {Conference} and {Ninth} {AAAI} {Symposium} on {Educational} {Advances} in {Artificial} {Intelligence}}, {AAAI}'19/{IAAI}'19/{EAAI}'19, pages 4213--4220, Honolulu, Hawaii, USA, 2019. AAAI Press.

\bibitem{rose1993constrained}
Kenneth Rose, Eitan Gurewitz, and Geoffrey~C Fox.
\newblock Constrained clustering as an optimization method.
\newblock {\em IEEE Transactions on Pattern Analysis and Machine Intelligence}, 15(8):785--794, 1993.

\bibitem{basu2008constrained}
Ian Wagstaff, Kiri Davidson, and Sugato Basu, editors.
\newblock {\em Constrained Clustering: {Advances} in Algorithms, Theory, and Applications}.
\newblock Chapman and Hall/CRC, New York, 2008.

\bibitem{goldstein2011vygotskys}
Vygotsky’s social development theory.
\newblock In Sam Goldstein and Jack~A. Naglieri, editors, {\em Encyclopedia of {Child} {Behavior} and {Development}}, pages 1550--1550. Springer US, Boston, MA, 2011.

\end{thebibliography}
\end{document}